# Ultralow Thermal Conductance of the van der Waals Interface between Organic Nanoribbons

Yucheng Xiong,[1,*] Xiaoxiang Yu[2,3,*], Yajie Huang[4], Juekuan Yang[5], Liangliang Li[4], Nuo Yang[2,3,†], and Dongyan Xu[1,‡]

[1]*Department of Mechanical and Automation Engineering, The Chinese University of Hong Kong, Shatin, New Territories, Hong Kong Special Administrative Region, People's Republic of China*

[2]*State Key Laboratory of Coal Combustion, Huazhong University of Science and Technology, Wuhan, 430074, People's Republic of China*

[3]*Nano Interface Center for Energy (NICE), School of Energy and Power Engineering, Huazhong University of Science and Technology, Wuhan 430074, People's Republic of China*

[4]*State Key Laboratory of New Ceramics and Fine Processing, School of Materials Science and Engineering, Tsinghua University, Beijing 100084, People's Republic of China*

[5]*School of Mechanical Engineering and Jiangsu Key Laboratory for Design and Manufacture of Micro-Nano Biomedical Instruments, Southeast University, Nanjing, 210096, People's Republic of China*

**Abstract**

Understanding thermal transport through nanoscale van der Waals interfaces is vital for addressing thermal management challenges in nanoelectronic devices. In this work, the interfacial thermal conductance ($G_{CA}$) between copper phthalocyanine (CuPc) nanoribbons is reported to be on the order of $10^5$ Wm$^{-2}$K$^{-1}$ at 300 K, which is over two orders of magnitude lower than the value predicted by molecular dynamics (MD) simulations for a perfectly smooth interface between two parallelly aligned CuPc nanoribbons. Further MD simulations and contact mechanics analysis reveal that surface roughness can significantly reduce the adhesion energy and effective contact area between CuPc nanoribbons, and thus result in an ultralow $G_{CA}$. In addition, the adhesion energy at the interface also depends on the stacking configuration of two CuPc nanoribbons, which may also contribute to the observed ultralow $G_{CA}$.

Nanostructures, such as nanotubes, nanowires, and nanoribbons, are typically assembled into large-area arrays to construct efficient electronic and photonic devices [1, 2]. Therefore, a large density of nanoscale van der Waals (vdW) interfaces exists in those devices. Recently, vdW heterostructures, formed by vertically stacking 2D materials, have also drawn extensive research interest for applications in ultrathin electronic devices and other functional devices [3-6]. The shrinking size and escalating integration density of transistors impose serious challenges for thermal management of electronic devices, especially for those with high-density interfaces. Understanding thermal transport through nanoscale vdW interfaces is crucial for addressing heat dissipation problems in those devices.

Interfacial thermal conductance is related to both materials in contact and interface properties including surface roughness, interfacial bonding and dislocations [7, 8]. The vdW interface between nanostructures is featured with restricted contact area and weak interactions. It has been demonstrated that weak adhesion and rough surfaces could lead to a significant reduction in the interfacial thermal conductance [7-11]. In the past decade, several experimental studies were conducted on thermal transport through nanoscale vdW interfaces. Yang and co-workers measured the contact thermal resistance between multiwalled carbon nanotubes (MWCNTs) and reported that the interfacial thermal conductance is proportional to the nanotube diameter [12], which is explained by the anisotropic thermal conductivity and long phonon mean free path along the *c*-axis of graphite. Hirotani *et al*. studied the thermal boundary conductance between one end of a carbon nanotube and an Au surface, and showed that the interfacial thermal conductance depends on the orientation of anisotropic carbon-based materials [13]. Zhou *et al*. reported that the interfacial thermal conductance of the nanosized contact between an indium arsenide (InAs) nanowire and a silicon nitride ($SiN_x$) substrate is two orders of magnitude lower than the value predicted by the diffuse mismatch model [14], which is attributed to the weak adhesion strength of vdW interactions. Pettes *et al.* showed that the contact thermal conductance between a bismuth telluride ($Bi_2Te_3$) nanoplate and platinum (Pt) electrodes is one to two orders of magnitude lower than the predicted value for an atomically smooth interface [15]. While these studies provide important experimental data for understanding thermal properties of nanoscale vdW interfaces, thermal transport through the vdW interface between organic nanostructures has rarely been probed.

Recently, metal phthalocyanines, a class of organic semiconductor materials, have attracted much attention due to their advantages of facile synthesis, low-cost availability, tunable properties and flexibility. Their optical and electrical properties have been extensively studied for applications in organic photovoltaic cells [16, 17], light-emitting diodes [18], infrared electroluminescent diodes [19] and field-effect transistors [20-22]. The low thermal conductivity of organic metal phthalocyanines and the interfacial thermal resistance are of major concerns for heat dissipation in these devices [8, 23]. Besides, the stacking structure of planar metal phthalocyanine molecules provides unique opportunities to explore how molecular orientation affects thermal transport through the vdW interface.

In this work, we experimentally investigated thermal transport through the vdW interface between copper phthalocyanine (CuPc) nanoribbons. Multiple thermal measurements were carefully carried out on segments cut from the same CuPc nanoribbon to determine the thermal resistance of the contact region of two segments by using a suspended thermal bridge method [24]. An interface heat transfer model was developed to extract the interfacial thermal conductance ($G_{CA}$) between CuPc nanoribbons. Ultralow interfacial thermal conductance on the order of $10^5$ $Wm^{-2}K^{-1}$ was observed for the planar contact between CuPc nanoribbons, which is three orders of magnitude lower than the $G_{CA}$ previously reported for the point contact between MWCNTs [12]. Molecular dynamics (MD) simulations and contact mechanics analysis were performed to elucidate fundamental mechanisms for the observed ultralow $G_{CA}$.

As shown in Fig. 1(a), a CuPc molecule has a planar structure, which is comprised of a central Cu atom surrounded by four pyrrole rings. A benzene ring is attached to each pyrrole ring. Two neighbouring pyrrole rings are connected by an N atom. CuPc molecules form a quasi-one-dimensional molecular column via the π-π interaction, while adjacent columns are bonded with each other by the vdW force. Figure 1(a) also depicts the herringbone stacking of CuPc molecules for β-phase CuPc. In this work, CuPc nanoribbons were synthesized via a physical vapour deposition method (see Sec. I in Ref. [25]). The crystalline structure of CuPc nanoribbons was characterized by high-resolution atomic force microscopy (HRAFM). The HRAFM image of one measured CuPc nanoribbon sample (C1) is given in Fig. 1(e), which clearly shows the crystalline orientation. The lattice constants determined from surface scanning profiles [Figs. 1(f) and 1(g)] are 1.98 nm and 0.48 nm for the *a*-axis and *b*-axis, respectively, confirming that CuPc nanoribbons studied in this work are β-phase [26].

To determine the thermal resistance of the contact region of two CuPc nanoribbons ($R_{CC}$), thermal resistances of CuPc segments of the non-contact region and the contact thermal resistance between CuPc nanoribbons and heat source/sink should be properly deducted from the measured total thermal resistance of two CuPc nanoribbons with a planar contact. To achieve this goal, a uniform CuPc nanoribbon with a length of tens of microns was cut into five segments. Two segments were carefully aligned to form a planar contact, bridging two membranes of a suspended device as shown in Fig. 2(a) for sample C1. Other three segments were transferred onto measurement devices with a gap of 2 μm, 4 μm, and 6 μm, respectively [25]. Thermal measurements of four samples were conducted separately by using the thermal bridge method. The measured total thermal resistance of two CuPc nanoribbons with a planar contact ($R_{tot\_C}$) consists of three parts: the contact thermal resistance between CuPc nanoribbons and two membranes ($R_{CM}$), thermal resistances of CuPc segments excluding the contact region ($R_{CuPc1}$ and $R_{CuPc2}$), and the thermal resistance of the contact region ($R_{CC}$). Then, the $R_{CC}$ can be determined by

$$R_{CC} = R_{tot\_C} - R_{CM} - (R_{CuPc1} + R_{CuPc2}) = R_{tot\_C} - R_{CM} - \frac{1}{\kappa wt} \times (L_1 + L_2), \quad (1)$$

where $L_1$ and $L_2$ are suspended lengths from the contact region to heat source and heat sink, respectively, $\kappa$, $w$, and $t$ are the thermal conductivity, width, and thickness of the CuPc nanoribbon. Similarly, for each single CuPc nanoribbon, the measured thermal resistance ($R_{tot\_S}$) can be written as

$$R_{tot\_S} = R_{CM} + \frac{1}{\kappa wt} \times L_S, \quad (2)$$

where $L_S$ is the suspended length of the single CuPc nanoribbon. As discussed in our previous study [24], the $R_{CM}$ can be assumed as a constant for measurements of three single CuPc nanoribbons with different suspended lengths and then $R_{tot\_S}$ should vary linearly with $L_S$. A linear relationship is indeed observed between the measured $R_{tot\_S}$ and $L_S$ [25], which verifies the above assumption. The geometry information of CuPc nanoribbons was measured by using the scanning electron microscopy (SEM) and atomic force microscopy (AFM). The $R_{CM}$ and $\kappa$ can be extracted simultaneously from the measured $R_{tot\_S}$ by the linear fitting [25]. Subsequently, the $R_{CC}$ can be determined. Figure 2(b) plots the obtained $R_{CC}$ as a function of temperature for five samples (C1-C5). The $R_{CC}$ varies among samples and the values are on the order of $10^7$ KW$^{-1}$.

For heat conduction through an interface between two quasi-one-dimensional nanostructures, heat flows along the nanostructure horizontally but through the interface vertically. To determine the $G_{CA}$, some approximations were commonly made by researchers [27-30]. For example, Zhong *et al.* [28] performed MD simulations on the interfacial thermal resistance between parallelly aligned carbon nanotubes with an overlap region. In their study, the interfacial thermal resistance was calculated by simplifying the overlap region as a planar interface between two coaxial nanotubes joined end to end. In Yang *et al.*'s work [30], the thermal resistance of the overlap region of two MWCNTs was treated as a contact thermal resistor connected in series with two MWCNT thermal resistors with the half of the overlap length. In these studies, thermal resistances of CNTs are relatively small compared to the interfacial thermal resistance and thus the aforementioned approximations might be reasonable. However, for thermal transport through the planar contact between CuPc nanoribbons with low thermal conductivity, a fin heat transfer model should be considered in the contact region to extract the $G_{CA}$.

An analytical model was developed in this work, which assumes one-dimensional heat conduction in each nanoribbon and a constant $G_{CA}$ for the interface between two nanoribbons. Figure 2(c) illustrates a schematic of two parallelly aligned nanoribbons with an overlap length $L_C$. Heat flows along each nanoribbon in the horizontal direction and through the interface in the vertical direction, as indicated by arrows in Fig. 2(c). Two nanoribbons are assumed to have the same width, thickness and thermal conductivity since they are cut from the same CuPc nanoribbon. The steady-state heat diffusion equations for top and bottom nanoribbons in the contact region can be written as (see Sec. III in Ref. [25])

$$\kappa \frac{d^2 T_T}{dx^2} wtdx - G_{CA}(T_T - T_B)wdx = 0, \tag{3a}$$

$$\kappa \frac{d^2 T_B}{dx^2} wtdx + G_{CA}(T_T - T_B)wdx = 0, \tag{3b}$$

where $T_T$ and $T_B$ denote the temperatures of top and bottom nanoribbons, respectively. These equations are similar to the two-temperature model [31-34] and the two-channel thermal transport model [35]. Eq. (3) can be solved by applying adiabatic and constant temperature boundary conditions at two ends of each nanoribbon. Note that all the heat will be conducted through the interface between two nanoribbons, heat rate ($q$) can be calculated by integrating the heat flux over the interface and can be derived as,

$$q = \frac{\gamma \kappa wt(e^{\gamma L_C} - 1)(T'_H - T'_S)}{1 + e^{\gamma L_C} + \gamma(e^{\gamma L_C} - 1)(L_1 + L_2 + L_C/2)}, \tag{4}$$

where $\gamma = \sqrt{2G_{CA}/\kappa t}$, $T'_H$ is the temperature at the joint of the top nanoribbon and the heating membrane, and $T'_S$ is the temperature at the joint of the bottom nanoribbon and the sensing membrane. Thus, $R_{CC}$ can be determined as

$$R_{CC} = \frac{T_T|_{-L_C/2} - T_B|_{L_C/2}}{q} = \frac{L_C}{2\kappa wt} + \frac{\gamma L_C (e^{\gamma L_C} + 1)}{2(e^{\gamma L_C} - 1)} \frac{1}{wL_C G_{CA}}. \tag{5}$$

The $G_{CA}$ can be determined from the experimentally measured $R_{CC}$ by solving Eq. (5).

The calculated temperature profile in the contact region clearly deviates from the linear distribution for the segment in the non-contact region for sample C1 [25]. In Yang et al.'s work [30], the thermal resistance of the overlap region of two MWCNTs was treated as a contact thermal resistor connected in series with two MWCNT thermal resistors with the half of the overlap length. By adopting this approximation approach, the thermal resistance of the contact region can be expressed as

$$R_{CC,\,app} = \frac{L_C}{\kappa wt} + \frac{1}{wL_C G_{CA}}. \tag{6}$$

Compared to the analytical model we derived, the approximation approach overestimates the first term at the right hand side of Eq. (5) by a factor of 2, which will lead to overestimation of $G_{CA}$ for CuPc nanoribbons with relatively large intrinsic thermal resistances [25].

The extracted $G_{CA}$ is shown in Fig. 2(d) for five measured samples (C1-C5). At room temperature, the values of $G_{CA}$ range from $1.8\times10^5$ to $6.5\times10^5$ Wm$^{-2}$K$^{-1}$. The relative uncertainty in $G_{CA}$ is estimated to be in the range of 48% to 120% for five samples (C1-C5) (see Sec. IV in Ref. [25]). The $G_{CA}$ we obtained for the vdW interface between CuPc nanoribbons is two to five orders of magnitude lower than the values for typical welded interfaces [7, 8]. The $G_{CA}$ results reported in the literature for other nanoscale vdW interfaces are also shown in Fig. 2(d) for comparison. Yang et al. reported a $G_{CA}$ on the order of $10^8$ Wm$^{-2}$K$^{-1}$ for the point contact between MWCNTs [12], as shown by blue dash lines in Fig. 2(d). The $G_{CA}$ between one end of a MWCNT and an Au surface (pink dash line) is determined to be $8.6\times10^7$-$2.2\times10^8$ Wm$^{-2}$K$^{-1}$ by Hirotani et al. [13]. Zhou et al. [14] reported that the $G_{CA}$ between an InAs nanowire and a SiN$_x$ substrate (green dash line) is $4.7\times10^6$-$2.5\times10^7$ Wm$^{-2}$K$^{-1}$. The contact thermal resistance per unit area for the interface between a Bi$_2$Te$_3$ nanoplate and Pt electrodes is in the range of $2.8\times10^{-7}$-$10^{-5}$ m$^2$KW$^{-1}$ [15], corresponding to a $G_{CA}$ of $10^5$-$3.5\times10^6$ Wm$^{-2}$K$^{-1}$ (black dash line). Compared to these studies, the $G_{CA}$ we obtained for the planar contact between CuPc nanoribbons is one to three orders of magnitude lower than the $G_{CA}$ values reported for the interfaces of MWCNT/MWCNT [12], MWCNT/Au [13], and InAs nanowire/SiN$_x$ [14], but very close to the results for the interface of Bi$_2$Te$_3$ nanoplate/Pt [15].

To elucidate the underlying mechanisms responsible for the observed ultralow $G_{CA}$, we performed atomistic MD simulations on thermal transport through the interface between CuPc nanoribbons (see Sec. V in Ref. [25]). The inset of Fig. 3(a) depicts the MD simulation system of two parallelly aligned CuPc nanoribbons forming a planar contact. The equilibrium distance between two CuPc nanoribbons is about 0.4 nm, which is close to the interlayer spacing in the *c*-axis of β-phase CuPc. Non-equilibrium MD method was applied to calculate the temperature profiles of top and bottom CuPc nanoribbons and the results are fitted fairly well by using the analytical model we derived, as shown in Fig. 3(a). The $G_{CA}$ determined by the MD simulation is $1.54 \times 10^8$ Wm$^{-2}$K$^{-1}$, which is comparable to the simulation result previously reported for vdW interfaces [36] but over two orders of magnitude higher than our experimental values (1.8-$6.5 \times 10^5$ Wm$^{-2}$K$^{-1}$).

It is worth noting that the surfaces of CuPc nanoribbons are assumed to be perfectly smooth in the MD simulation, however, the samples we measured always have a certain level of surface roughness. The surface roughness of CuPc nanoribbons was characterized by using a contact mode AFM (AR Cypher, Oxford, UK). The scan size is 30 nm × 30 nm with 256 pixels for each axis. Figure 1(c) shows the AFM image of sample C1 processed with flattening and plane fitting routines. Figure 1(d) plots the surface height histogram corresponding to the image in Fig. 1(c), which can be fitted well by using a Gaussian distribution with a mean height of -0.03 nm and a root-mean-square (rms) roughness ($\sigma$) of 0.45 nm. Since the contact between two rough surfaces only occurs at peaks, the spacing between two surfaces varies from point to point, which will affect the adhesion energy between two surfaces. We first calculated the adhesion energy (see Sec. VI in Ref. [25]) between two perfectly smooth CuPc nanoribbons ($E$) when one ribbon is fixed and the other one is gradually moved away from the equilibrium position (the displacement or separation distance is denoted as $s$), as plotted in Fig. 3(b). The adhesion energy reaches the maximum value of 0.12 Jm$^{-2}$ ($E_0$) at the equilibrium position ($s=0$), which is a typical value for vdW interfaces [37]. The calculated $E$ decreases with $s$ quickly and approaches to zero when $s$ is larger than 8 Å. In contact mechanics, it is well accepted that the contact between two elastic rough surfaces with a rms roughness of $\sigma$ can be modeled as the contact between a flat elastic surface and a rigid rough surface with a rms roughness of $\sqrt{2}\sigma$ [38]. We followed the same approach in this work and treated two rough CuPc nanoribbons as a flat elastic surface and an equivalent rigid rough surface in contact. The mean separation distance between two contacting surfaces depends on material properties, surface profiles, and external load. At a given separation distance, the average adhesion energy ($\bar{E}$) can be calculated based on the variation of $E$ with $s$ for two perfectly smooth surfaces [green solid line in Fig. 3(b)] and the Gaussian distribution of surface heights for the equivalent rough surface. In the calculation, for the peaks with a surface height larger than the given separation distance, the corresponding adhesion energy is assumed to be $E_0$, which is similar to Maugis' approximation [39]. The variation of the calculated average adhesion energy with the mean separation distance is also given as a black solid line in Fig. 3(b). Notably, when the equivalent rough surface is squeezed onto the flat surface and mean planes of two surfaces are overlapped ($s=0$), the average adhesion energy is 0.065 Jm$^{-2}$, only 55% of the adhesion energy between two perfectly smooth CuPc nanoribbons. In our experiments, no external force is applied on two CuPc nanoribbons, so the mean separation distance should be larger than zero. In the literature, simple approaches have been proposed to estimate the mean separation distance between two surfaces from surface roughness. For example, Rumpf *et al.* suggested to take the mean separation distance as 1.485 times of the rms roughness for the contact between a smooth particle and a rough surface [40]. This gives a mean separation distance of 0.95 nm and a $\bar{E}/E_0$ of 13%, as indicated by the red dot in Fig. 3(b). Rabinovich *et al.* suggested that the mean separation distance could be approximated by 1.817 times of the rms surface roughness [41], corresponding to a $\bar{E}/E_0$ of 7% [blue dot in Fig. 3(b)]. According to Prasher's model [11], $G_{CA}$ is proportional to the square of the surface adhesion strength. By using the mean separation distances proposed by Rumpf *et al.* and Rabinovich *et al.*, the $G_{CA}$ between two rough CuPc nanoribbons can be estimated from the MD result for two perfectly smooth CuPc nanoribbons and are equal to $2.6 \times 10^6$ Wm$^{-2}$K$^{-1}$ and $7.55 \times 10^5$ Wm$^{-2}$K$^{-1}$, respectively, which are close to our experimental results.

The above analysis shows that surface roughness will lead to an increase of the mean separation distance between two CuPc nanoribbons. As a result, the adhesion energy and the interfacial thermal conductance between two rough CuPc nanoribbons could be substantially lower than the counterparts for two perfectly smooth CuPc nanoribbons in the MD simulation. The contact between two rough CuPc nanoribbons may be reexamined in terms of the effective contact area by taking adhesion into account. Due to their elastic nature, adhesion will occur for two contacting CuPc nanoribbons. At molecular scale, the vdW interaction pulls two surfaces into contact, which will decay quickly with the increase of the distance. When two CuPc nanoribbons form a planar contact, a certain portion of surfaces is too far away from each other to sustain the adequate adhesion due to the surface roughness. The real contact area of an adhesive contact strongly depends on adhesion strength, material properties, and roughness parameters [42, 43]. Surface roughness is taken into account through the dimensionless rms slope $h' \equiv \langle |\nabla h|^2 \rangle^{1/2}$, where $h$ is the height profile of the rough surface. $h'$ is estimated to be 0.63 for CuPc nanoribbons according to the AFM image [Fig. 1(c)]. Recently, Persson and Scarraggi showed that the relative contact area $A/A_0$ is around 0.06 at zero external load for surfaces with adhesion energy of 0.1 Jm$^{-2}$, rms roughness of 0.6 nm, and $h'$ of 0.0035 [42]. CuPc nanoribbons measured in this work have adhesion energy (0.12 Jm$^{-2}$) and rms roughness (0.45 nm) similar to the surface studied in Ref. [42] but much larger $h'$ (0.63). It should be noted that surface heights can only be measured at discrete points by AFM and experimental noise becomes a critical issue at small scan sizes. Indeed, experimental results of $h'$ could vary over a wide range depending on the measurement technique and the scan size. It is still under debate how to measure $h'$ accurately [44, 45]. Instead of pursuing its precise value, we intend to discuss the role of $h'$ in determining $A/A_0$ for adhesive contacts. CuPc nanoribbons synthesized via the physical vapor deposition method possess highly crystalline structure and rather clean surfaces, as evidenced by AFM and SEM. We expect that $h'$ will be smaller than the estimated value (0.63). McGhee *et al*. experimentally showed that $A/A_0$ decreases with the increase of $h'$ [46]. Therefore, $A/A_0$ for CuPc nanoribbons is expected to be smaller than the value (0.06) predicted by Persson and Scaraggi [42]. Considering the effective contact area, the $G_{CA}$ between two rough CuPc nanoribbons is estimated to be lower than 9.24 × 10$^6$ Wm$^{-2}$K$^{-1}$.

It should be noted that there are several possible stacking configurations for two CuPc nanoribbons. Considering that molecular stacking may affect the adhesion between two contacting surfaces, we calculated adhesion energies for different stacking configurations of two CuPc nanoribbons using MD simulations. Figure 3(c) enumerates four stacking configurations: pristine stacking, *a*-axis translocation, bottom up and *ab*-plane rotation. In the pristine stacking, the atomic arrangement at the interface is identical to the packing inside a CuPc nanoribbon. On the basis of the pristine stacking, one ribbon is turned upside down in the bottom up configuration. In the *a*-axis translocation, CuPc molecules in one nanoribbon are slid by half of the lattice constant along the *a*-axis, while one nanoribbon is rotated by 90° for the *ab*-plane rotation. As seen in Fig. 3(d), the pristine stacking demonstrates the highest adhesion energy (0.12 Jm$^{-2}$) among four stacking configurations, while the adhesion energy of the *ab*-plane rotation is only 0.07 Jm$^{-2}$, corresponding to 58% of the pristine stacking. In view of the fact that the stacking configuration of CuPc nanoribbons in our experiments is similar to the *ab*-plane rotation [Fig. 2(a)], the lower adhesion energy of this configuration may also contribute to the observed ultralow $G_{CA}$.

Furthermore, some other surface phenomena such as surface reconstruction may occur in nanostructures [47] and will also affect interfacial thermal transport between CuPc nanoribbons. Unfortunately, similar to many other potential functions [48], the force field used in this work is unable to accurately describe surface reconstruction of CuPc nanoribbons. Thus, the impact of surface reconstruction on interfacial thermal transport is not taken into account in our MD simulations. Previous studies suggested that surface reconstruction would decrease surface energies of Si, Ge and Au [48-50]. Compared to Si or Au nanomaterials, CuPc nanoribbons should have a much lower surface energy due to weak intermolecular interactions. Besides, as seen from the HRAFM image [Fig. 1(e)], the surface of CuPc nanoribbons exhibits very good lattice ordering. Therefore, we expect surface reconstruction may not be significant for CuPc nanoribbons.

This work sheds light on understanding thermal transport through the vdW interface between nanostructures. Distinct from the point contact between MWCNTs, surface roughness plays a pronounced role for thermal transport through the planar contact between CuPc nanoribbons. MD simulations and contact mechanics analysis reveal that surface roughness will significantly reduce the adhesion energy and the effective contact area between CuPc nanoribbons, which will result in orders of magnitude lower $G_{\text{CA}}$. This explains the ultralow $G_{\text{CA}}$ observed for CuPc nanoribbons as well as other planar contact [15]. In addition, our MD simulations disclose that the adhesion energy at the interface depends on the stacking configuration, which may also contribute to the observed ultralow $G_{\text{CA}}$.

D.X. acknowledges the funding support from the Research Grants Council of the Hong Kong Special Administrative Region, People's Republic of China, under the General Research Fund (RGC Ref. No. 14238416). N.Y. acknowledges the funding support from National Natural Science Foundation of China (Grant No. 51576076 and No. 51711540031), Natural Science Foundation of Hubei Province (Grant No. 2017CFA046) and Fundamental Research Funds for the Central Universities (HUST: Grant No. 2016YXZD006). J.Y. acknowledges the funding support from the National Natural Science Foundation of China (Grant No. 51676036). L.L. acknowledges the funding support from the National Natural Science Foundation of China (Grant No. 51572149). The authors thank the National Supercomputing Center in Tianjin (TianHe-1 (A)) and China Scientific Computing Grid (ScGrid) for providing assistance in computations.

[†]nuo@hust.edu.cn

[‡]dyxu@mae.cuhk.edu.hk

[*]Y.X. and X.Y. contributed equally to this work.

**Figures**

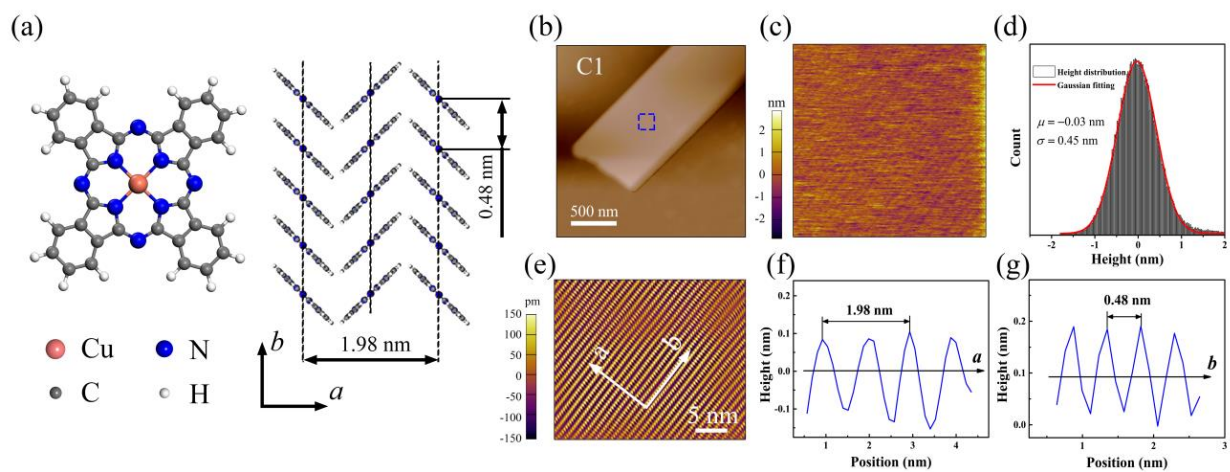

FIG. 1. Molecular structure and AFM characterization of CuPc nanoribbons. (a) Schematics of the planar CuPc molecule and the herringbone stacking of CuPc molecules for β-phase CuPc. (b) AFM image of one measured CuPc nanoribbon sample (C1). The blue dash square indicates where HRAFM was taken. (c) HRAFM image of sample C1 processed with flattening and plane fitting routines. The scan size is 30 nm × 30 nm. (d) The surface height histogram for the image in (c) and the Gaussian fitting curve. (e) HRAFM image of sample C1 processed with flattening, plane fitting and Fourier transform. (f, g) Surface scanning profiles for the *a*-axis and *b*-axis. The lattice constants determined for the *a*-axis and *b*-axis are 1.98 nm and 0.48 nm, respectively, confirming that CuPc nanoribbons are β-phase.

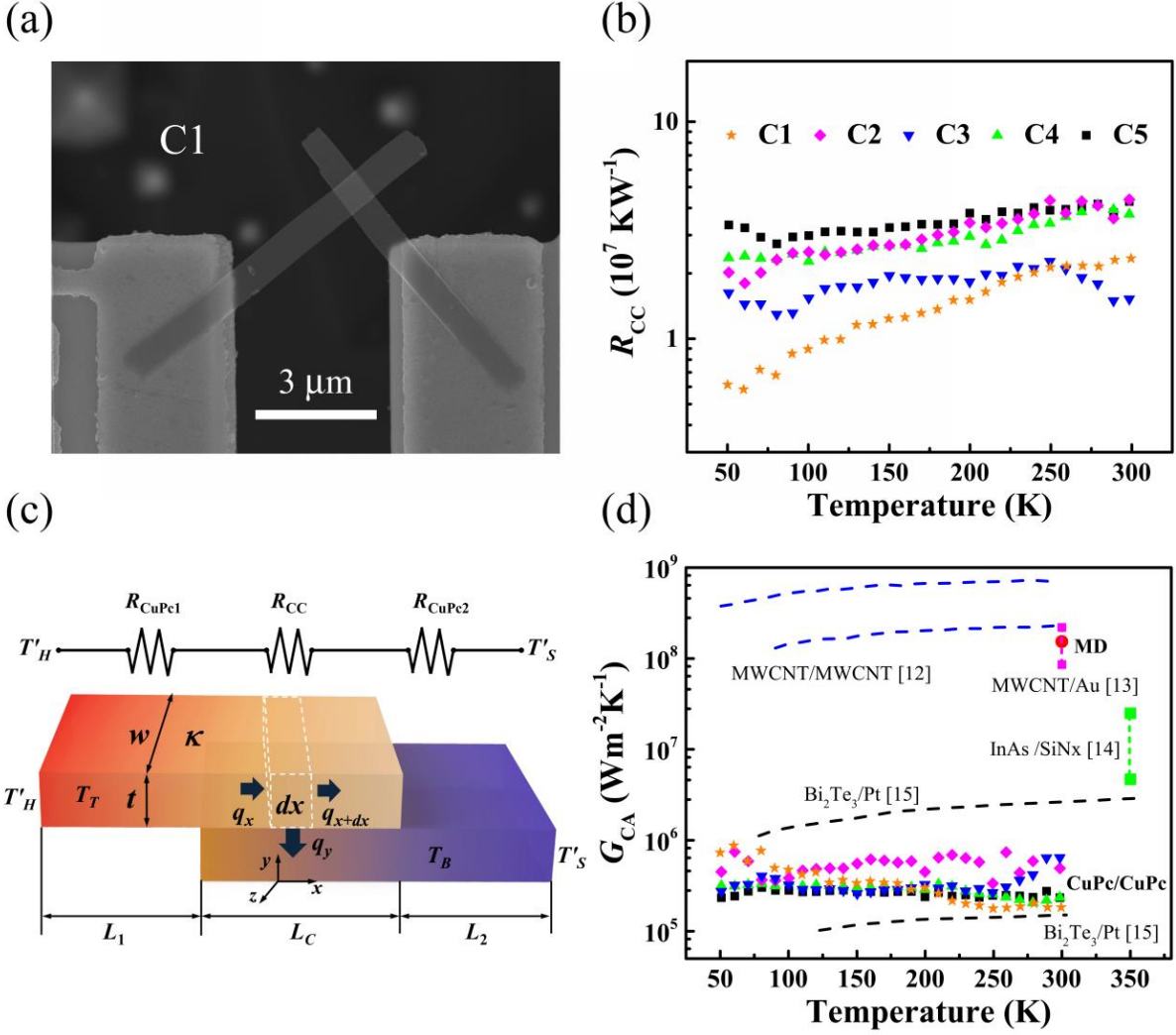

FIG. 2. Interfacial thermal conductance for the vdW interface between CuPc nanoribbons. (a) SEM micrograph of two CuPc nanoribbons with a planar contact (C1) on a suspended device for thermal measurement. (b) Experimentally determined $R_{CC}$ for five samples (C1-C5). (c) Schematic of two nanoribbons in contact with an overlap length $L_C$ used for deriving the interface heat transfer model and the corresponding thermal circuit. (d) Experimentally determined $G_{CA}$ and the MD simulation result. The $G_{CA}$ results reported for other vdW interfaces in the literature are also shown for comparison.

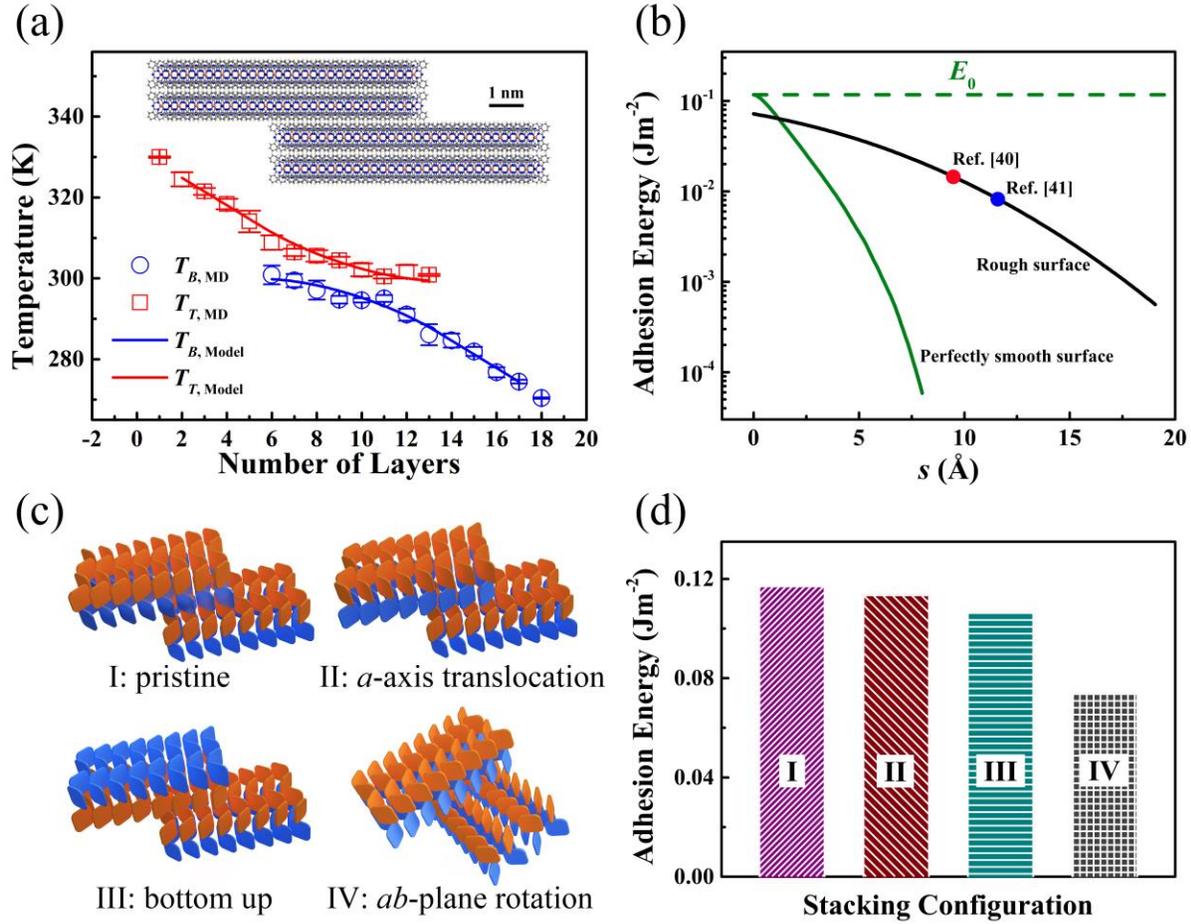

FIG. 3. MD simulations on thermal transport through the interface between CuPc nanoribbons. (a) Temperature profiles of top and bottom nanoribbons calculated by the MD simulation (symbols) and the fitting curves by using the interface heat transfer model (solid lines). The inset shows the MD simulation box of two parallelly aligned CuPc nanoribbons forming a planar contact. The molecular stacking at the interface is identical to that inside the CuPc nanoribbon (pristine stacking). (b) Adhesion energies calculated for perfectly smooth (green solid line) and rough CuPc nanoribbons (black solid line) as a function of the separation distance. The green dash line shows the adhesion energy for two perfectly smooth CuPc nanoribbons at the equilibrium position ($s=0$) as the reference. Red and blue dots denote average adhesion energies estimated for two rough CuPc nanoribbons by using the mean separation distances

proposed in Ref. [40] and [41], respectively. (c) Schematics of four stacking configurations. (d) Calculated adhesion energies for different stacking configurations.

# Supplemental Material

### I. Materials Synthesis

CuPc nanoribbons were synthesized via a physical vapour deposition method, as described in our previous study [S1]. CuPc powders (>99.95%, Sigma-Aldrich) were placed at the high-temperature zone of a horizontal three-zone tube furnace, and vaporized at 450°C for an hour. The vapour was carried by high-purity argon gas at a rate of 200 sccm from the high-temperature zone to the low-temperature zone. Single crystalline CuPc nanoribbons were formed on a silicon substrate placed at the low-temperature zone (200-300°C). The synthesized CuPc nanoribbons are β-phase with a growth direction of [010].

### II. Measured Samples

A total of five samples (C1-C5) have been measured in this work. Figure S1 shows SEM images of three single CuPc nanoribbons and two CuPc nanoribbons with a planar contact, and measured thermal resistances for five samples (C1-C5).

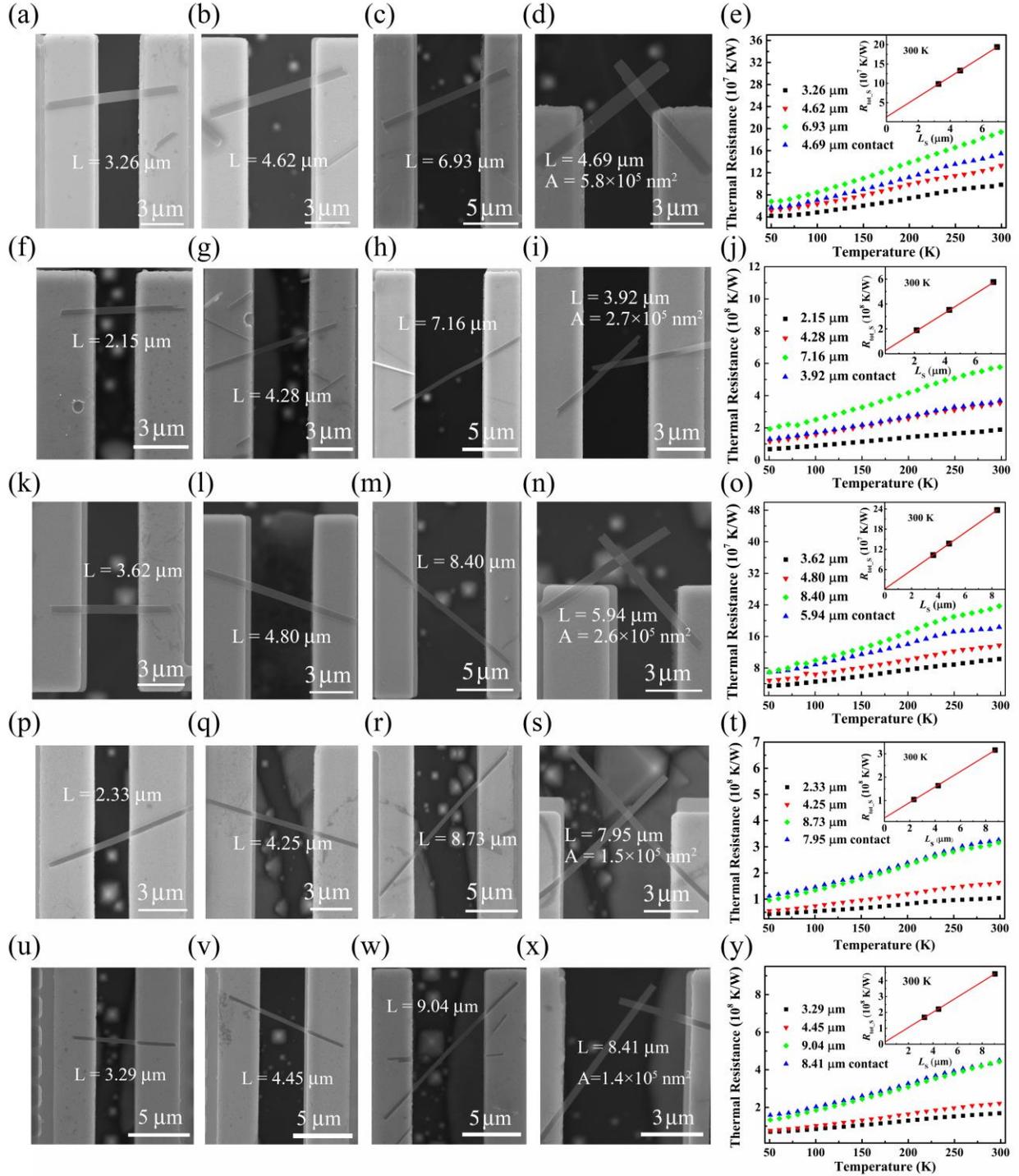

FIG. S1. SEM images and measured thermal resistances for five samples (C1-C5). (a-e) for C1; (f-j) for C2; (k-o) for C3; (p-t) for C4; and (u-y) for C5. The insets of (e, j, o, t and y) show measured thermal resistances of single CuPc nanoribbons versus suspended lengths at 300 K. The thickness is 113 nm, 69 nm, 104 nm, 85 nm and 97 nm for C1-C5, respectively.

Figure S2 plots the extracted contact thermal resistance between the CuPc nanoribbon and two membranes ($R_{CM}$) and thermal conductivity ($\kappa$) for five samples (C1-C5).

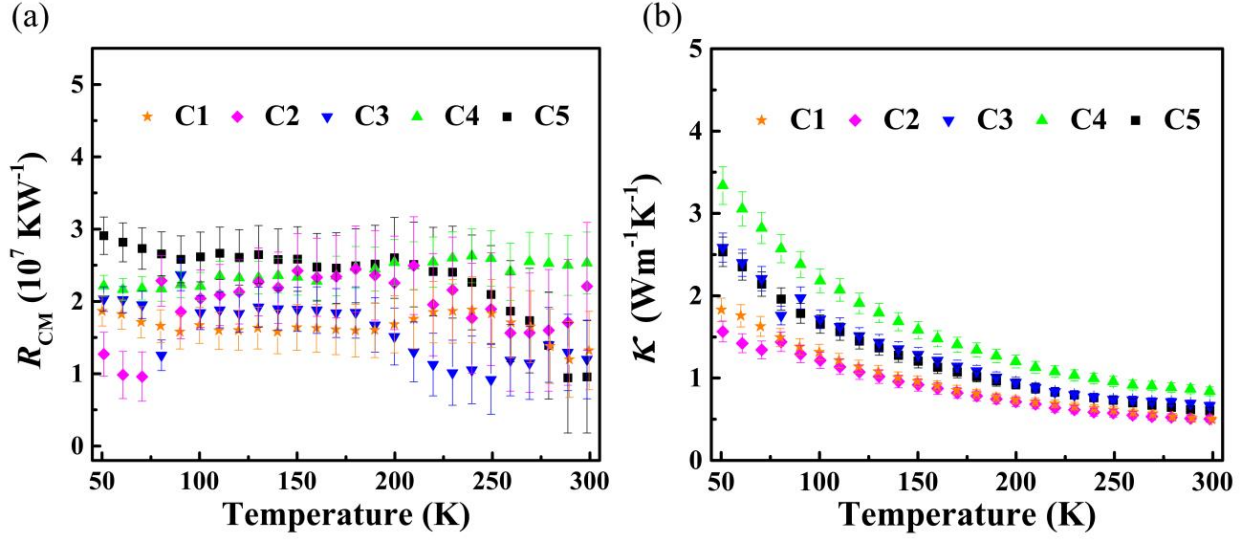

FIG. S2. Extracted contact thermal resistance (a) and thermal conductivity (b) for five samples (C1-C5).

### III. Interface Heat Transfer Model

An interface heat transfer model is derived to extract the interfacial thermal conductance between CuPc nanoribbons ($G_{CA}$). As shown in the schematic in Fig. 2(c), the model considers a configuration of two nanoribbons forming an aligned contact with an overlap length $L_C$. The model assumes one-dimensional heat conduction in each nanoribbon and a constant $G_{CA}$ between two nanoribbons. Under the steady-state condition, for a differential control volume selected in the top nanoribbon in the contact region ($-L_C/2 \leq x \leq L_C/2$), heat rates at the left, right, and bottom surfaces can be expressed as

$$q_x = -\kappa w t \frac{dT_T}{dx}, \quad \text{[S1a]}$$

$$q_{x+dx} = q_x + \frac{dq_x}{dx}dx = q_x - \kappa w t \frac{d^2 T_T}{dx^2} dx, \quad \text{[S1b]}$$

$$q_y = G_{CA}(T_T - T_B) w dx, \quad \text{[S1c]}$$

where $T_T$ and $T_B$ denote the temperatures of top and bottom nanoribbons, respectively, $w$ and $t$ are the width and thickness of the nanoribbon. The top, front, and back surfaces of the differential control volume are assumed to be adiabatic. Applying the conservation of energy principle to the differential control volume, a heat diffusion equation can be obtained for the top nanoribbon,

$$\kappa \frac{d^2 T_T}{dx^2} wtdx - G_{CA}(T_T - T_B)wdx = 0, \quad -\frac{L_C}{2} \leq x \leq \frac{L_C}{2}. \qquad [\text{S2a}]$$

Similarly, the following equation can be derived for the bottom nanoribbon,

$$\kappa \frac{d^2 T_B}{dx^2} wtdx + G_{CA}(T_T - T_B)wdx = 0, \quad -\frac{L_C}{2} \leq x \leq \frac{L_C}{2}. \qquad [\text{S2b}]$$

Subtracting Eq. S2b from Eq. S2a and assuming $\theta_1 = T_T - T_B$, we obtain

$$\frac{d^2 \theta_1}{dx^2} - \gamma^2 \theta_1 = 0, \quad -\frac{L_C}{2} \leq x \leq \frac{L_C}{2}, \qquad [\text{S3}]$$

where

$$\gamma = \sqrt{2 G_{CA} / \kappa t}. \qquad [\text{S4}]$$

The general solution of Eq. S3 is

$$\theta_1 = Ae^{\gamma x} + Be^{-\gamma x}, \quad -\frac{L_C}{2} \leq x \leq \frac{L_C}{2}, \qquad [\text{S5}]$$

where $A$ and $B$ are unknown constants. Combining Eq. S2a with Eq. S2b and assuming $\theta_2 = T_T + T_B$, we obtain

$$\frac{d^2 \theta_2}{dx^2} = 0, \quad -\frac{L_C}{2} \leq x \leq \frac{L_C}{2}. \qquad [\text{S6}]$$

The general solution of Eq. S6 is of the form

$$\theta_2 = Cx + D, \quad -\frac{L_C}{2} \leq x \leq \frac{L_C}{2}, \qquad [\text{S7}]$$

where $C$ and $D$ are unknown constants. Thus, temperature distributions of top and bottom nanoribbons in the contact region ($-L_C/2 \leq x \leq L_C/2$) are

$$T_T = \frac{\theta_1 + \theta_2}{2} = \frac{1}{2}(Ae^{\gamma x} + Be^{-\gamma x} + Cx + D),  \quad \text{[S8a]}$$

$$T_B = \frac{\theta_2 - \theta_1}{2} = \frac{1}{2}(-Ae^{\gamma x} - Be^{-\gamma x} + Cx + D). \quad \text{[S8b]}$$

In the non-contact region, heat diffusion equations for top and bottom nanoribbons are

$$\kappa w t \frac{d^2 T_T}{dx^2} dx = 0, \quad -L_1 - \frac{L_C}{2} \leq x < -\frac{L_C}{2}, \quad \text{[S9a]}$$

$$\kappa w t \frac{d^2 T_B}{dx^2} dx = 0, \quad \frac{L_C}{2} < x \leq L_2 + \frac{L_C}{2}, \quad \text{[S9b]}$$

where $L_1$ and $L_2$ are suspended lengths from the contact region to heat source and heat sink, respectively.

Adiabatic and constant temperature boundary conditions are assumed for each nanoribbon,

$$T_T \big|_{-L_1 - \frac{L_C}{2}} = T'_H \text{ and } \frac{dT_T}{dx}\bigg|_{\frac{L_C}{2}} = 0, \quad \text{[S10a]}$$

$$\frac{dT_B}{dx}\bigg|_{-\frac{L_C}{2}} = 0 \text{ and } T_B \big|_{L_2 + \frac{L_C}{2}} = T'_S, \quad \text{[S10b]}$$

where $T'_H$ is the temperature at the joint of the top nanoribbon and the heating membrane, and $T'_S$ is the temperature at the joint of the bottom nanoribbon and the sensing membrane. Applying the boundary conditions in Eq. S10 to Eq. S8 and Eq. S9, the constants can be determined as

$$A = B = \frac{[T'_H - T'_S - q(L_1 + L_2)/\kappa w t]e^{\gamma L_C/2}}{1 + e^{\gamma L_C} + (e^{\gamma L_C} - 1)\gamma L_C/2}, \quad \text{[S11a]}$$

$$C = \frac{\gamma [T'_H - T'_S - q(L_1 + L_2)/\kappa w t](1 - e^{\gamma L_C})}{1 + e^{\gamma L_C} + (e^{\gamma L_C} - 1)\gamma L_C/2}, \quad \text{[S11b]}$$

$$D = T'_H + T'_S - q(L_1 - L_2)/\kappa w t, \quad \text{[S11c]}$$

where $q$ is the heat rate along the nanoribbon.

Temperature distributions of the top and bottom nanoribbons are solved as

$$T_T = \begin{cases} \dfrac{q}{\kappa wt}\left(x+L_1+\dfrac{L_C}{2}\right)+T'_H, & -L_1-\dfrac{L_C}{2}\le x < -\dfrac{L_C}{2} \\[6pt] \dfrac{1}{2}\left\{\dfrac{T'_H-T'_S-\dfrac{q(L_1+L_2)}{\kappa wt}}{1+e^{\gamma L_C}+(e^{\gamma L_C}-1)\dfrac{\gamma L_C}{2}}\left[e^{\dfrac{\gamma L_C}{2}}(e^{\gamma x}+e^{-\gamma x})+\gamma(1-e^{\gamma L_C})x\right]+T'_H+T'_S-\dfrac{q(L_1-L_2)}{\kappa wt}\right\}, & -\dfrac{L_C}{2}\le x \le \dfrac{L_C}{2} \end{cases}$$ [S12a]

$$T_B = \begin{cases} \dfrac{1}{2}\left\{\dfrac{T'_H-T'_S-\dfrac{q(L_1+L_2)}{\kappa wt}}{1+e^{\gamma L_C}+(e^{\gamma L_C}-1)\dfrac{\gamma L_C}{2}}\left[-e^{\dfrac{\gamma L_C}{2}}(e^{\gamma x}+e^{-\gamma x})+\gamma(1-e^{\gamma L_C})x\right]+T'_H+T'_S-\dfrac{q(L_1-L_2)}{\kappa wt}\right\}, & -\dfrac{L_C}{2}\le x \le \dfrac{L_C}{2} \\[6pt] \dfrac{q}{\kappa wt}\left(x-L_2-\dfrac{L_C}{2}\right)+T'_S, & \dfrac{L_C}{2} < x \le \dfrac{L_C}{2}+L_2 \end{cases}$$ [S12b]

Note that all the heat will be conducted through the interface between two nanoribbons. $q$ can be calculated by integrating the heat flux over the interface,

$$q = \int_{-L_C/2}^{L_C/2} G_{CA}\theta_1 w dx = \gamma\kappa wt(e^{\gamma L_C}-1)\dfrac{T'_H-T'_S-q(L_1+L_2)/\kappa wt}{1+e^{\gamma L_C}+(e^{\gamma L_C}-1)\gamma L_C/2}.$$ [S13a]

Thus, $q$ can be determined as

$$q = \dfrac{\gamma\kappa wt(e^{\gamma L_C}-1)(T'_H-T'_S)}{1+e^{\gamma L_C}+\gamma(e^{\gamma L_C}-1)(L_1+L_2+L_C/2)}.$$ [S13b]

The thermal resistance of the contact region between two nanoribbons can be calculated by

$$R_{CC} = \dfrac{T_T|_{-L_C/2}-T_B|_{L_C/2}}{q} = \dfrac{e^{\gamma L_C}+1+(e^{\gamma L_C}-1)\gamma L_C/2}{\gamma\kappa wt(e^{\gamma L_C}-1)} = \dfrac{L_C}{2\kappa wt}+\dfrac{\gamma L_C(e^{\gamma L_C}+1)}{2(e^{\gamma L_C}-1)}\dfrac{1}{wL_C G_{CA}}$$
$$= R_1+\dfrac{\gamma L_C(e^{\gamma L_C}+1)}{2(e^{\gamma L_C}-1)}R_2$$ [S14]

Notably, $R_1 = L_C/2\kappa wt$ stands for the conduction thermal resistance of two parallel nanoribbons, and $R_2 = 1/wL_C G_{CA}$ represents the interface thermal resistance between two nanoribbons. Figure S3 plots the calculated temperature profiles of top and bottom CuPc nanoribbons for sample C1. It is clearly shown that temperature distributions of both top and bottom nanoribbons in the contact region deviate from the linear distribution in the non-contact region.

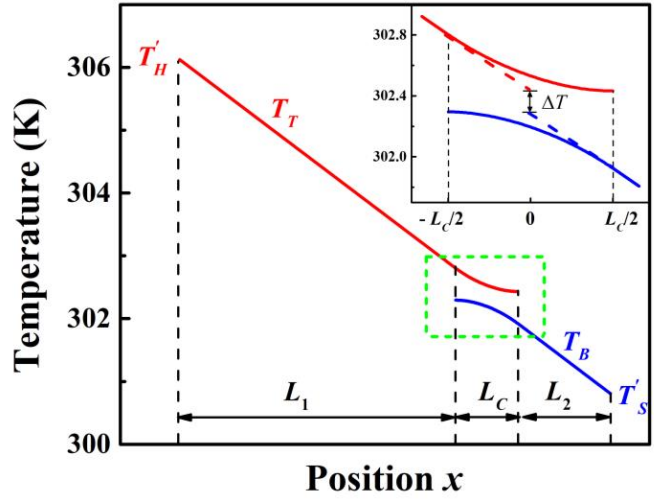

FIG. S3. The calculated temperautre profiles of top and bottom CuPc nanoribbons for sample C1.

In Yang *et al.*'s work [S2], the thermal resistance of the overlap region of two MWCNTs was treated as a contact thermal resistor connected in series with two MWCNT thermal resistors with the half of the overlap length. By adopting this approximation approach, the corresponding temperature profiles are shown as dash lines in the inset of Fig. S3. Thus, the thermal resistance of the contact region can be approximated by

$$R_{CC, app} = \frac{L_C}{\kappa wt} + \frac{1}{wL_C G_{CA}} = 2R_1 + R_2. \qquad [S15]$$

Compared to Eq. S14 we derived, the approximation approach overestimates the first term by a factor of 2 and assumes that the coefficient in front of $R_2$ is equal to 1. For the further comparison, Eq. S14 can be rewritten as

$$\begin{aligned}R_{CC} &= 2R_1 + \left[\frac{\gamma L_C(e^{\gamma L_C}+1)}{2(e^{\gamma L_C}-1)} - \frac{R_1}{R_2}\right]R_2 = 2R_1 + \left[\frac{\gamma L_C(e^{\gamma L_C}+1)}{2(e^{\gamma L_C}-1)} - \left(\frac{\gamma L_C}{2}\right)^2\right]R_2 \\ &= 2R_1 + \left[\frac{\lambda}{\tanh(\lambda)} - \lambda^2\right]R_2\end{aligned} \qquad [S16]$$

where $\lambda = \gamma L_C/2 = \sqrt{R_1/R_2}$. The only difference between Eq. S15 and Eq. S16 is the coefficient in front of $R_2$. Figure S4 plots the coefficient $\lambda/\tanh(\lambda) - \lambda^2$ as a function of $\lambda$. When $\lambda = 0$ (corresponding to $R_1 \ll R_2$), the coefficient is equal to 1, indicating that the analytical model we derived coincides with the approximation approach. However, for CuPc nanoribbons with relatively large intrinsic thermal resistances, $\lambda$ is always larger than zero. Thus, the coefficient will be less than 1 as seen in Fig. S4, indicating that the approximation approach will overestimate the $G_{CA}$.

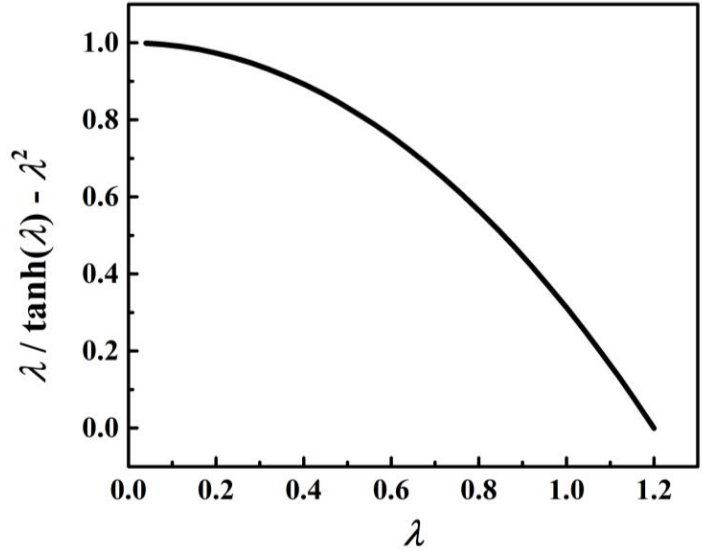

FIG. S4. Coefficient $\lambda/\tanh(\lambda) - \lambda^2$ as a function of $\lambda$.

**IV. Uncertainty Analysis**

The per unit length thermal resistance of the CuPc nanoribbon ($R'_{CuPc}$) and the contact thermal resistance ($R_{CM}$) are determined from the measured thermal resistances of three single nanoribbons by the linear least squares fitting:

$$R'_{CuPc} = \frac{n\sum_{i=1}^{n} L_{S,i} R_{tot\_S,i} - \sum_{i=1}^{n} L_{S,i} \sum_{i=1}^{n} R_{tot\_S,i}}{n\sum_{i=1}^{n}(L_{S,i}^2) - \left(\sum_{i=1}^{n} L_{S,i}\right)^2}, \quad [S17a]$$

$$R_{CM} = \frac{\sum_{i=1}^{n}(L_{S,i}^2)\sum_{i=1}^{n} R_{tot\_S,i} - \sum_{i=1}^{n} L_{S,i} \sum_{i=1}^{n}(L_{S,i} R_{tot\_S,i})}{n\sum_{i=1}^{n}(L_{S,i}^2) - \left(\sum_{i=1}^{n} L_{S,i}\right)^2}, \quad [S17b]$$

where $L_{S,i}$ and $R_{tot\_S,i}$ represent the suspended length and the measured thermal resistance of the *i*th nanoribbon, $n$ is the number of single nanoribbons measured for each sample, which is 3 in this study. The relative uncertainties in $R'_{CuPc}$ and $R_{CM}$ are estimated by following the standard approach of uncertainty propagation [S3],

$$\frac{\delta R'_{CuPc}}{R'_{CuPc}} = \sqrt{\sum_{i=1}^{n}\left(\frac{\partial R'_{CuPc}}{\partial R_{tot\_S,i}}\right)^2 \left(\frac{\delta R_{tot\_S,i}}{R'_{CuPc}}\right)^2 + \sum_{i=1}^{n}\left(\frac{\partial R'_{CuPc}}{\partial L_{S,i}}\right)^2 \left(\frac{\delta L_{S,i}}{R'_{CuPc}}\right)^2}, \qquad [S18a]$$

$$\frac{\delta R_{CM}}{R_{CM}} = \sqrt{\sum_{i=1}^{n}\left(\frac{\partial R_{CM}}{\partial R_{tot\_S,i}}\right)^2 \left(\frac{\delta R_{tot\_S,i}}{R_{CM}}\right)^2 + \sum_{i=1}^{n}\left(\frac{\partial R_{CM}}{\partial L_{S,i}}\right)^2 \left(\frac{\delta L_{S,i}}{R_{CM}}\right)^2}, \qquad [S18b]$$

where $\delta R_{tot\_S,i}$ and $\delta L_{S,i}$ are uncertainties in the measured thermal resistance and the suspended length, respectively. The relative uncertainty in the measured thermal resistance ($\delta R_{tot\_S,i}/R_{tot\_S,i}$) is evaluated by using the Monte Carlo method and is less than 4% in the temperature range of 50-100 K, and ~2% above 100 K. The suspended length is measured from the SEM image, and the error $\delta L_{S,i}$ is estimated to be 10 nm. Thus, the relative uncertainty in the thermal conductivity can be calculated by

$$\frac{\delta\kappa}{\kappa} = \sqrt{\left(\frac{\delta R'_{CuPc}}{R'_{CuPc}}\right)^2 + \left(\frac{\delta w}{w}\right)^2 + \left(\frac{\delta t}{t}\right)^2}, \qquad [S19]$$

where $\delta w$ and $\delta t$ are uncertainties in the width and thickness of the nanoribbon, respectively. The width is determined from the SEM image and the uncertainty is estimated as 2 nm. The thickness is measured by atomic force microscopy, and the uncertainty is estimated as 5 nm.

For two nanoribbons with a planar contact, the thermal resistance of the contact region is determined by

$$R_{CC} = R_{tot\_C} - R_{CM} - R'_{CuPc} \times (L_1 + L_2). \qquad [S20]$$

The relative uncertainty in the $R_{CC}$ is estimated by

$$\frac{\delta R_{CC}}{R_{CC}} = \sqrt{\left(\frac{\delta R_{tot\_C}}{R_{CC}}\right)^2 + \left(\frac{\delta R_{CM}}{R_{CC}}\right)^2 + \left(R'_{CuPc}\frac{\delta L_1}{R_{CC}}\right)^2 + \left(R'_{CuPc}\frac{\delta L_2}{R_{CC}}\right)^2 + \left[(L_1+L_2)\frac{\delta R'_{CuPc}}{R_{CC}}\right]^2}. \qquad [S21]$$

The $G_{CA}$ can be determined by solving Eq. S14. The Monte Carlo method [S3] is adopted to estimate the uncertainty in the derived $G_{CA}$. An appropriate probability distribution function is assumed for each error source, for which Gaussian and rectangular distribution functions are chosen for random errors and systemtic errors, respectively. The simulation was run for 1000 times to obtain the distribution of $G_{CA}$, from which the standard deviation is estimated. The uncertainty in $G_{CA}$ at a 95% level of confidence is two times the standard deviation. At room temperature, the estimated relative uncertainty in $G_{CA}$ ranges from 48% to 120% for five samples (C1-C5).

## V. Molecular Dynamics (MD) Simulations

MD simulations were performed using a large-scale atomic/molecular massively parallel simulator (LAMMPS) package. Class II force field potential forms including high-order functions and cross terms were applied to describe interatomic interactions including structural components (bonds, angles, and dihedrals) and nonbonding interactions (Lennard-Jones and Coulomb interactions). All the potential parameters were taken from Ref. [S4]. A cut-off distance of 10 Å was used for both the Lennard-Jones and Coulomb potentials. Boundaries along *a*, *b* and *c* axes were set to be free, fixed and free, respectively. A time step of 0.5 fs was chosen due to fast vibrating hydrogen atoms. The details of the LAMMPS package are listed in Table SI for CuPc nanoribbons.

Table SI. The details of the LAMMPS package for CuPc nanoribbons

| Method | Non-equilibrium MD (Direct method) | | | | |
|---|---|---|---|---|---|
| **Potentials** | | | | | |
| Style | Bond style class2  Angle style class2  Dihedral style class2  Improper style class2  Pair_style lj/class2 10.0 & coul/dsf 0.4 10.0  Mixing rule sixthpower  Special_bonds lj 0.0 0.0 0.5 coul 0.0 0.0 0.5  Kspace_style none | | | | |
| Parameters | Ref. [S4] | | | | |
| **Simulation System** | | | | | |
| Unit cell: Ref. [S4]  Sizes of nanoribbons: 7.5 nm (length) × 4 nm (width) × 2.5 nm (thickness) | | | | | |
| **Setting** | | | | | |
| Time step | 0.5 fs | | Pressure | | 0 atm |
| Ensembles | Simulation process | | | | Purpose |
| NVT | Temperature | 300 K | Run time | 0.5 ns | Relax the structure |
| NVE | Heat source | 330 K | Heat sink | 270 K | Establish a steady-state temperature gradient |
| | Run time | 2 ns | Thermostat | Langevin | |
| NVE | Heat source | 330 K | Heat sink | 270 K | Record information |
| | Run time | 3 ns | Thermostat | Langevin | |
| **Recorded physical quantity** | | | | | |
| **Temperature** | $<E> = \sum_{i=1}^{N} \frac{1}{2} m v_i^2 = \frac{1}{2} N k_B T_{\text{MD}}$ | | | | |

As depicted in the inset of Fig. 3(a), the MD simulation system consists of two CuPc nanoribbons forming a planar contact. The positions of CuPc molecules at two ends of the simulation system are fixed. The molecules adjacent to the fixed ones are set as heat source and heat sink, respectively, using Langevin thermostats. Temperature profiles are recorded from the MD simulation after reaching steady state and are shown in Fig. 3(a). As seen in Fig. S5, the tallied energies of heat sink and heat source show a linear trend, which verifies that the system reaches steady state.

By performing a linear fitting for the cumulative energy of heat source/sink, we obtain a heat rate of $2.015 \times 10^{-8}$ W for the simulated system. With the known cross-sectional area and temperature gradient, the thermal conductivity of the CuPc nanoribbon is calculated to be 0.19 Wm$^{-1}$K$^{-1}$ at 300 K. The thermal conductivity obtained from the MD simulation is much lower than experimental values [Fig. S2(b)] due to the size effect. Temperature profiles obtained from the MD simulation are fitted fairly well by using the analytical solution in Eq. S12. The $G_{CA}$ determined by the MD simulation is $1.54 \times 10^8$ Wm$^{-2}$K$^{-1}$.

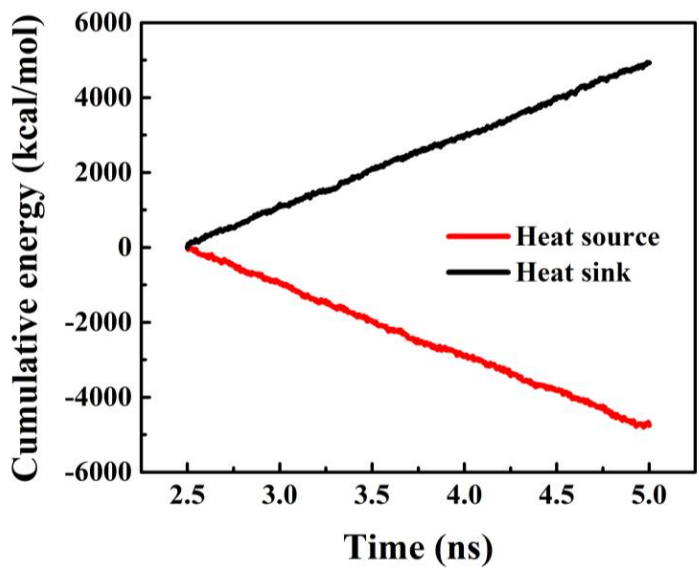

FIG. S5. Cumulative energies of heat baths in the MD simulation at 300 K.

## VI. Calculation of the Adhesion Energy

To determine the adhesion energy between two CuPc nanoribbons, we simulate a process of separating two nanoribbons from the equilibrium distance. The system consisting of two nanoribbons in contact is first equilibrated for 100 ps to reach the equilibrium position. Then, all atoms of one CuPc nanoribbon are fixed, while the other nanoribbon is gradually moved away up to 20 Å. The change in the potential energy for the whole system is recorded, which corresponds to the adhesion energy between two CuPc nanoribbons. The calculated adhesion energy as a function of the displacement is shown as a green solid line in Fig. 3(b).